\documentclass[longauth]{aa}  

\usepackage{pythonhighlight}

\usepackage{color}
\usepackage[colorlinks,breaklinks]{hyperref}
\hypersetup{linkcolor=blue,citecolor=blue,filecolor=black,urlcolor=blue}
\usepackage{amssymb}

\newcommand{\slobject}[1]{\texttt{#1}}

\usepackage{xspace}

\usepackage{graphicx}
\usepackage{txfonts}
\usepackage{orcidlink}

\usepackage[utf8]{inputenc}

\usepackage[switch]{lineno}

\begin{document}

 \title{skysurvey: a pure python package to simulate the transient sky.}
   \author{
    Rigault, M.\inst{\ref{ip2i}}\fnmsep\thanks{\texttt{m.rigault@ip2i.in2p3.fr}} \orcidlink{0000-0002-8121-2560},
    Ginolin, M.\inst{\ref{cambridge}} \orcidlink{0009-0004-5311-9301},
    Dellazzeri, L.\inst{\ref{ip2i}}\orcidlink{0009-0006-8378-0111},
    Popovic, B.\inst{\ref{southampton}}\orcidlink{0000-0002-8012-6978},
    Osman Hjortlund, J.\inst{\ref{okc}}, 
    Gilles Lordet, A. \inst{\ref{okc}}\orcidlink{0009-0002-4464-8063}, 
    Conseil, S.\inst{\ref{ip2i}}, 
    Coughlin, M. \inst{\ref{uminessota}},
    Ruppin, F. \inst{\ref{ip2i}},
    Smith, M. \inst{\ref{lancaster}}\orcidlink{0000-0002-8012-6978},
    Townsend, A. \inst{\ref{birmingham}}\orcidlink{0000-0001-6343-3362},
    Trigui, A.\inst{\ref{ip2i}}\orcidlink{0009-0005-9981-8121},
    Barjou-Delayre, C.\inst{\ref{clermont}}\orcidlink{0009-0000-8510-8982},
    Kebadian, R. \inst{\ref{cppm}}\orcidlink{0009-0003-4093-1870},
    Nordin, J. \inst{\ref{berlin}}\orcidlink{0000-0001-8342-6274},
    }   
   \institute{
   Université Lyon 1, CNRS, IP2I Lyon, UMR 5822, Villeurbanne, France \label{ip2i} 
   \and
   Institute of Astronomy and Kavli Institute for Cosmology, Madingley Road, Cambridge CB3 0HA, UK \label{cambridge}
   \and
   School of Physics and Astronomy, University of Southampton, Southampton, SO17 1BJ, UK
   \label{southampton}
   \and
   The Oskar Klein Centre, Department of Physics, AlbaNova, SE-106 91 Stockholm , Sweden
   \label{okc}
   \and   
   School of Physics and Astronomy, University of Minnesota, Minneapolis, Minnesota 55455, USA
   \label{uminessota}
   \and
   Department of Physics, Lancaster University, Lancs LA1 4YB, UK 
   \label{lancaster}
   \and
    School of Physics and Astronomy, University of Birmingham, Birmingham B15 2TT, UK \label{birmingham}
    \and
     Université Clermont Auvergne, CNRS, LPCA, F-63000 Clermont-Ferrand, France
    \label{clermont} 
   \and
   Aix Marseille Univ, CNRS/IN2P3, CPPM, Marseille, France \label{cppm}
   \and
   Institut fur Physik, Humboldt-Universit\"{a}t zu Berlin, 12489 Berlin, Germany
   \label{berlin}
   }
\titlerunning{\texttt{skysurvey}}
\authorrunning{M. Rigault et al.}

\abstract
{Accurate simulation of astronomical observations is a critical element for any modern analyses, be it to measure event rates, analyses population properties, validate or train pipelines, account for selection effects, or correct biases. }
{We present a novel pure python package named \texttt{skysurvey} made to enable the user to quickly simulate astrophysical transients as observed by a survey. The package is structured to make the implementation of any complex population modeling fast and easy.}
{\texttt{skysurvey} relies on three core objects: a \texttt{Target}, that models how an astrophysical target exists in nature, a \texttt{Survey}, that specifies how the sky has been observed and, a \texttt{DataSet} that combine these two to generate data as they would have been acquired. In addition, we present a side stand-alone package named \texttt{modeldag} that contains the core structure that simplifies the parameter modeling.}
{We present in this paper how \texttt{skysurvey} is structured and we clearly illustrate how the code can straightforwardly be used to simulate complex populations, such as Type Ia Supernovae with varying color-brightness $\beta$ term. We also illustrate how the package can be made to replicate the rate and redshift distribution of the ZTF SNe Ia DR2 dataset.}
{The \texttt{skysurvey} package, already used in recent scientific publications, is now ready for general usage and paves the way for future use of simulations such as simulation based inference.}


   \maketitle

\section{Introduction}

Simulations are the foundation of many scientific analyzes, as they allow scientists to validate methods on cases where the ground truth is known. This is particularly useful when dealing with selection effects that impact the vast majority of astronomical and cosmological studies ; be it because  of instrumental limits, because the Milky Way, or, say, of the availability of follow-up instruments. Simulations are then used to validate inference pipelines or to probe what has not been observed given what has. The latter aspect is key to derive rates \citep[e.g.][]{pain2002,leaman2011,perley2020}, to study properties of the underlying population \citep[e.g.][]{li2011,vincenzi2019,dimitriadis2025} or to correct the selection effect when deriving cosmological parameters \citep[e.g.,][]{betoule2014,kessler2017,scolnic2018,riess2019,brout2022}. The use of massive realistic simulations has furthermore been shown to be able to train machine-learning methods to infer model parameters, a technique now known as simulation based inference \citep[e.g.][and references therein]{cranmer2020,karchev2023,Ho2024, boyd2026}. Finally, simulations are essential to plan future instruments and surveys \citep[e.g.][]{goobar2002,kessler2010,bernstein2012,nissanke2013,hounsell2018}. 

In extra-galactic transient sciences, the \texttt{SNANA} package \citep{kessler2009} has long been the central simulation software. 
Among many other key successes, \texttt{SNANA} has been the reference software used to prepare the upcoming Rubin Large Survey of Space and Time \citep[LSST,][]{ivezic2019} and the cosmology program of the Roman Space Telescope \citep{rose2021,rose2025}. There, \texttt{SNANA} was most notably the software suite used to simulate realistic transients as part of the PLAsTiCC and ELAsTiCC classification challenges \citep{kessler2019, hlozek2023}, which have been extensively used to develop and explore classification algorithms. Outside its use for transients modeling, \texttt{SNANA} has been the central tool behind the last decade of precision cosmological measurements from Type Ia Supernovae (SNe~Ia) \citep[e.g.][]{scolnic2018, brout2022, vincenzi2024, popovic2025} via the Beams with Bias Correction method \citep[BBC,][]{scolnic2016,kessler2017,popovic2021}. 

But \texttt{SNANA} is not without downsides, mostly stemming from its long legacy. \texttt{SNANA} is a complex combination of \texttt{C}, \texttt{fortran} and \texttt{python}, making installation famously complex and largely limited to a handful of high performance computing servers. Further compounding the approachability problem is the legacy of its 10+ years of development: the software contains hard-coded numbers and structures that predate original analyses approach to recently discovered issues. Notably, the SN-host correlations \citep[e.g.,][]{sullivan2010,rigault2013,roman2018,kelsey2021,rigault2020,broutscolnic2021} was not considered in the initial development of \texttt{SNANA}. 
This led recent analyses to  iterative fudging and new implementations of the SN-to-host matching to recover observed distributions for the treatment of SN-to-host correlation across, e.g., \cite{smith2020}, \cite{broutscolnic2021}, \cite{popovic2021}, \cite{popovic2023}, and \cite{wiseman2022}. Because of this long development cycle, these new features are piped through pre-defined structures such as libraries of host galaxies (\texttt{HOSTLIB}s) and hard-coded probability distribution functions (\texttt{WGTMAPs}).
\texttt{SNANA} uses \texttt{.input} files, ideal for long-term storage and replicability, but complex and spread across multiple fronts within \texttt{SNANA}. The addition of the \texttt{pippin} wrapper \citep[][]{hinton2020} has simplified the user experience, but, crucially, has not fixed the issue of IO and \texttt{SNANA}'s inability to rapidly iterate over different model, SN, or survey properties that would be necessary for the implementation of, say, SBI.

In light of this situation, alternative simulation codes have emerged to replace the simulation layer of \texttt{SNANA}. So far, most focused on specific cases, such as \texttt{simsurvey} \citep{feindt2019} to prepare the Zwicky Transient Facility survey \citep[ZTF][]{bellm2019} and then extensively used by the collaboration to derive rates \citep[e.g.,][]{sagues2021,andreoni2022,karambelkar2023}, or \texttt{snsim}  \citep{carreres2023} developed to prepare the large-scale structure inference by nearby Type Ia Supernovae (SNe Ia). In 2022, we have started the development of \texttt{skysurvey} with the aim to replace the simulation layer of \texttt{SNANA}. The goal of this package is to provide a fast running \texttt{pip install}able library that would enable any user to model (complex) simulations of any kind of astrophysical sources, and notably of any transients, as given by nature and as observed by any kind of survey (e.g., ZTF or LSST). As this paper was being written, the \texttt{LightCurveLynx} survey simulation package has been introduced through a first example on simulated SN Ia populations \citep{dai2026}, further illustrating the need for alternative simulation codes to \texttt{SNANA}.

This paper presents \texttt{skysurvey} version 1.0. Ongoing or future developments shall continue improving the code, both in terms of speed and flexibility, while conserving the core aspect of the package: an easy to use software made to enable the user to simulate complex astrophysical targets and surveys. We invite the user to refer to the \href{https://skysurvey.readthedocs.io/en/latest}{online documentation} that contains many usage examples including those described in this paper. The code is publicly available on \href{https://github.com/MickaelRigault/skysurvey}{GitHub} and can be installed using the \texttt{pip install skysurvey} command.

Since its initial development, the \texttt{skysurvey} package has already been used to publish scientific analyses: to reproduce an existing dataset to understand its characteristics \citep{amenouche2025, rigault2025}, to simulate realistic Supernovae data to validate an inference pipeline \citep{ginolin2025a, ginolin2025b}, to study the origin of unexpected rate variations with redshift \citep{gilles2026}, to compare observed Supernovae rate with expectations in early LSST data \citep{freeburn2025}, to analyse the observations of strongly lensed supernovae \citep{sagues2024}, to forecast measurements of $H_0$ anisotropies with ZTF \citep{barjou2026}, etc. 
More analyses are ongoing, such as a complex Type Ia Supernovae modeling (Ginolin et al. in prep), a validation of dark energy and large scale structure cosmological inference pipelines (Betoule et al. in prep, Kebadian et al. in prep.). The \texttt{skysurvey} package is also particularly well suited to train simulation based inference pipeline (Trigui et al. in prep), to derive core-collapse Supernovae rate and luminosity functions (Qin et al. in prep) or to validate strongly lensed supernovae analyses (Osman Hjortlun et al. in prep.).

The paper presents \texttt{skysurvey}, focusing on introducing its core concept, structure, and functionalities. 
We start by presenting "\texttt{modeldag}" in Section~\ref{sec:modeldag}, a standalone package built in parallel of \texttt{skysurvey} and that provides its basis to draw target parameters, making it easy to build complex astrophysical models. Then, after a rapid overview of the \texttt{skysurvey} threefold structure (target, survey and dataset) in Section~\ref{sec:structure} we introduce each with more details: \texttt{Target}s in Section~\ref{sec:target}, \texttt{Survey}s in Section~\ref{sec:survey} and \texttt{DataSet} in Section~\ref{sec:dataset}. We continue by illustrating a simple test example in Section~\ref{sec:example} where we reproduce the redshift distribution of the ZTF SN Ia DR2 sample \citep{rigault2025}. We conclude in Section~\ref{sec:conclusion}.

Throughout the paper, we present code-lines in figures in parallel to their physical discussion in the main text. This may be unusual in scientific publication, but we are convinced it is the best way to illustrate the fundamental idea behind \texttt{skysurvey}: A package made to respect the logical flow of a(n) (astro)physics intuition, which in turns enables the user to easily build complex (realistic) simulations thanks a straightforward user interface. 

\section{\texttt{modeldag}: build any parameter model}
\label{sec:modeldag}

At the core of \texttt{skysurvey}'s parameter generation is the \texttt{modeldag}\footnote{\url{https://modeldag.readthedocs.io/en/latest/}} package. This \texttt{python} package enables a user to build complex models from a simple dictionary structure, forming a Directed Acyclic Graph (DAG). In this DAG, each node represents a model parameter, and the edges represent dependencies between them. Each parameter is associated to a function that describes how this parameter is drawn. Thanks to the DAG structure, the output of one parameter draw can be used as input of the function that draws another. The strength of \texttt{modeldag} is that this process is made straightforward by its simple \texttt{python} dictionary structure.

This is fully illustrated in Fig.~\ref{fig:modeldag} for a simple model inspired by the SNe~Ia brighter-slower brighter-bluer relations \citep[e.g.,][]{tripp1998,riess1996}: $m_i=\alpha\times x_{1,i} + \beta \times c_i$, with $\alpha$ and $\beta$ population parameters and $m_i$, $x_{1,i}$ and $c_i$ individual SNe~Ia terms. As shown in the middle part of the left panel in Fig.~\ref{fig:modeldag}, The \texttt{modeldag} (nested-) dictionary structure is the following. The first level \texttt{keys} ($x1$, $c$ and $m$, Fig.~\ref{fig:modeldag}) represent the parameters to be generated. The second level explains how to do so, using only 2 entries: "\texttt{func}", which provide the information of the function to be used to generate the parameter, and "\texttt{kwargs}", which provide the function's parameters. 
The core idea behind \texttt{modeldag} is that one can use the symbol ``@``  to use the result of a parameter's drawing as parameter input for another parameter, as illustrated in Fig.~\ref{fig:modeldag}. In that example, $x1$ and $c$ are drawn from normal distribution, but $m$ is generated by combining them linearly ; notice the \texttt{x1='@x1'} and \texttt{c='@c'} in $m$'s \texttt{kwargs}. 

\subsection{Any function to generate parameters}
\label{sec:modelfunc}

\begin{figure*}
  \sidecaption
  \includegraphics[width=0.7\linewidth]{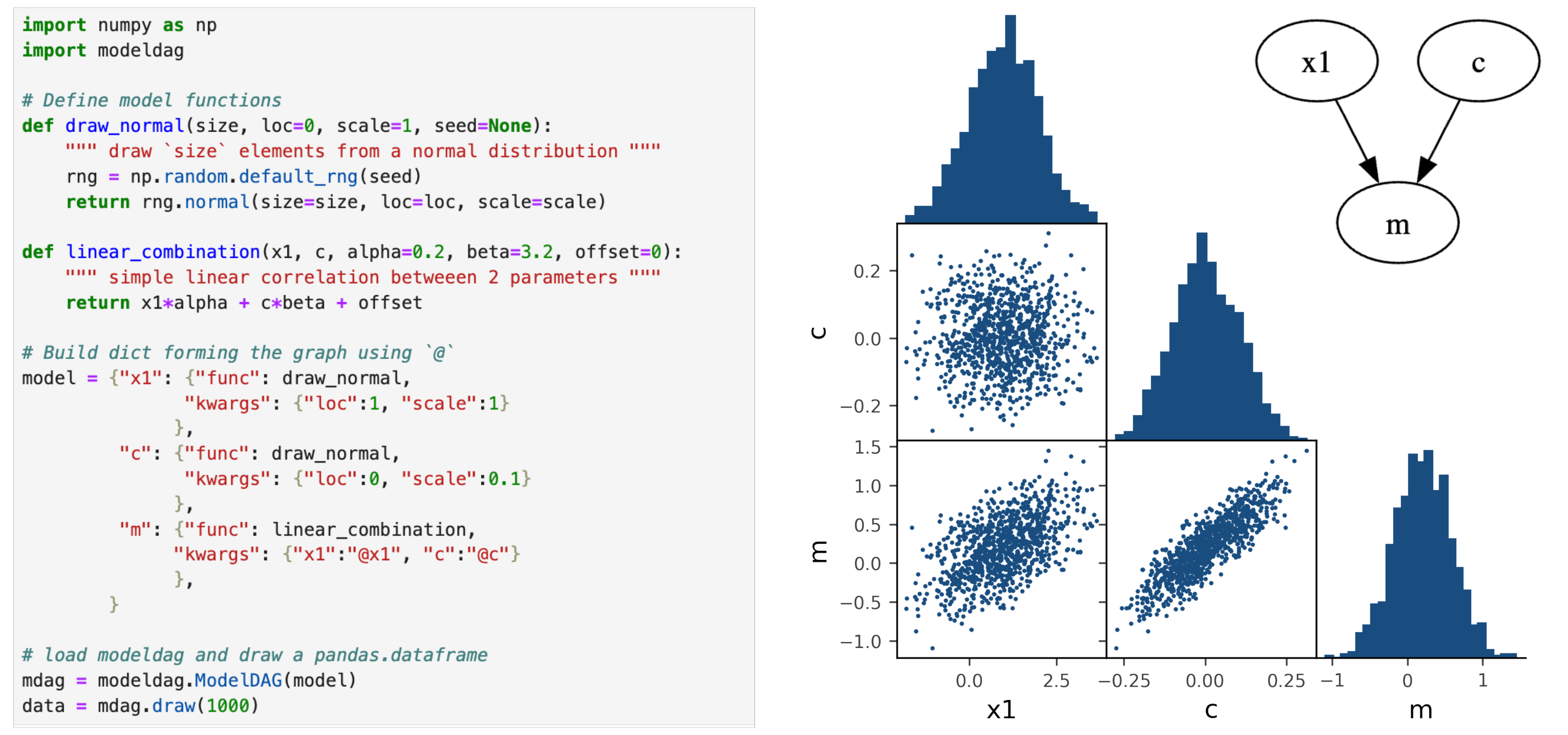}
  \caption{Illustration of \texttt{modeldag}. 
  \emph{left}: a \texttt{python} code generating a simple  model, where two randomly drawn parameters ($x1$ and $c$) are linearly combined to form a third ($m$). 
  The "$@$" sign enables to pass parameters from a key to another to form a graph as illustrated in the \emph{top-right} panel. 
  \emph{right:} Corner plot presenting the generated data.
  }
  \label{fig:modeldag}
\end{figure*}

Any function can be used to generate parameters.
As illustrated in Fig.~\ref{fig:modeldag} \texttt{modeldag} handles functions that return parameter(s) or that "transform" them (i.e. generate one from others). To handle functions returning multiple parameters generated simultaneously, the second level dictionary accepts a third keyword along side \texttt{func} and \texttt{kwargs}: "\texttt{as}". This last keyword enables the user to specify each returned parameter name. As illustrated in Fig.~\ref{fig:modeldag_extension}, this is useful when several parameters have to be generated simultaneously, as when randomly drawing points uniformly distributed in a shell like sky-coordinates (RA, Dec). Alternatively, \texttt{as} could be used to rename a single parameter.

In addition, \texttt{modeldag} accepts functions that return (n-d) probability distribution functions (PDF). To be automatically identified by \texttt{modeldag}, these functions must return two arrays: '\texttt{xx}' and '\texttt{pdf}'. Then \texttt{modeldag} will sample the parameter by randomly drawing \texttt{xx} values following the individual point probability given by \texttt{pdf} ; see online documentation for examples.

Across the code, the \texttt{size} argument must be accepted by functions that returned drawn parameters, like \texttt{draw\_normal} in Fig.~\ref{fig:modeldag}. Transformation functions do not need it as \texttt{size} will be taken from the length of input parameters (like $x1$ or $c$, Fig.~\ref{fig:modeldag}). Functions returning PDF should accept \texttt{xx} as input argument specifying where the pdf must be estimated ; input \texttt{xx} and returned \texttt{xx} do not need to match.

\subsection{Complexifying a model, automatic graph}
\label{sec:modelcomplex}

\begin{figure}
  \includegraphics[width=0.7\linewidth]{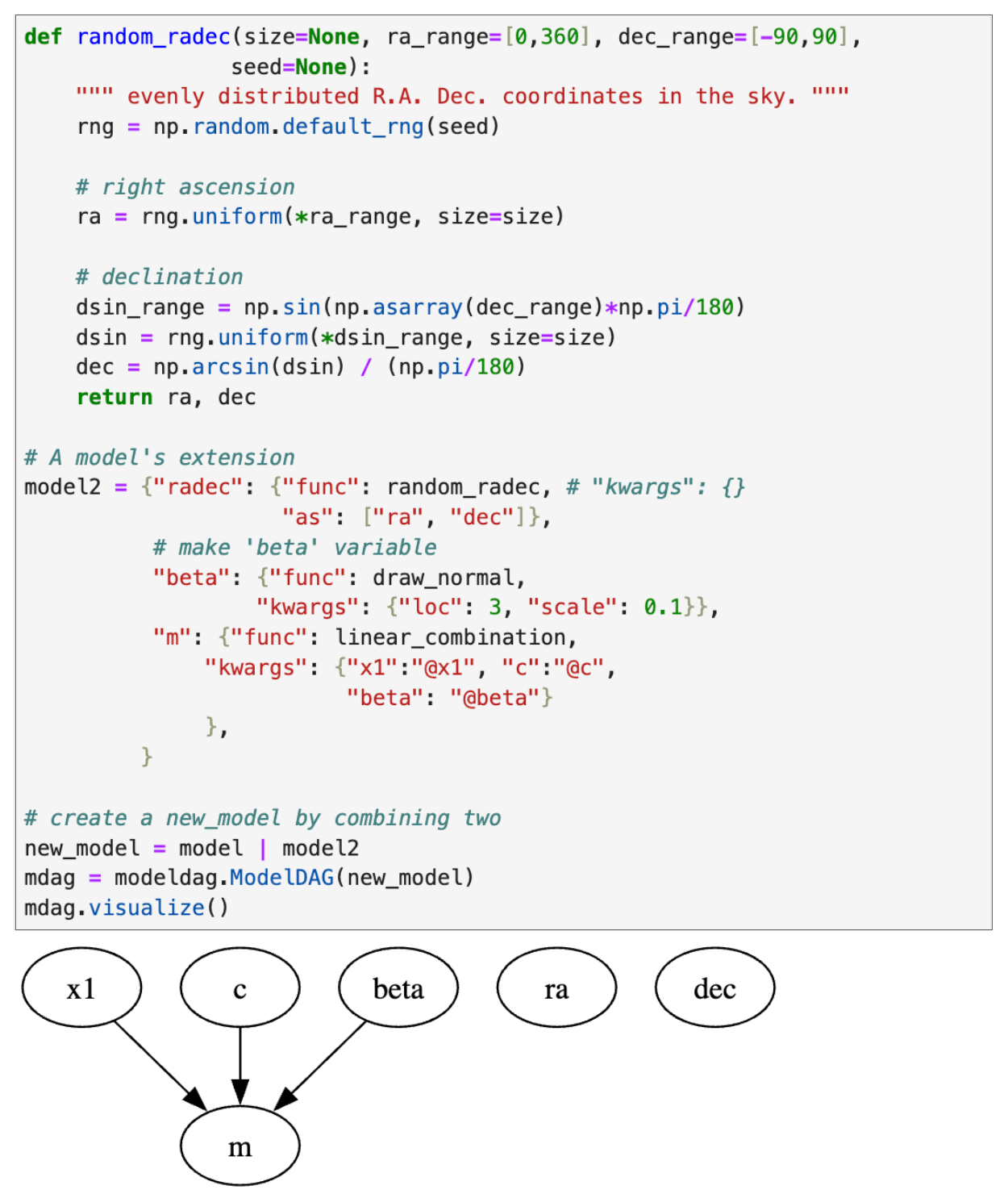}
  \caption{Complexifying a \texttt{modeldag}'s model. 
  \emph{Top}: a \texttt{python} code creating a model, including a function that returns two variables (\texttt{ra} and \texttt{dec}) generated simultaneously). There a new model is generated by combining \texttt{model\_2} with the original one shown in Fig.~\ref{fig:modeldag}. 
  \emph{bottom:} new graph.
  }
  \centering
  \label{fig:modeldag_extension}
\end{figure}

Thanks to \texttt{modeldag}'s "@" symbol trick, it is easy to build logical connections between parameters. Furthermore, since the overall structure is a dictionary, it benefits from all \texttt{python} dictionary tools. Updating a \texttt{model}'s entry or adding new parameters and/or dependency is as simply as updating a python's dictionary. This is illustrated in Fig.~\ref{fig:modeldag_extension}. In that example, the \texttt{ra} and \texttt{dec} coordinates are generated simultaneously and a new \texttt{beta} parameter is introduced. The parameter is then used as input for the \texttt{m}'s transformation function (see Fig.~\ref{fig:modeldag}). By combining the initial \texttt{model} from Fig.~\ref{fig:modeldag} and that extension one can have a more complex model (called \texttt{new\_model}, Fig.~\ref{fig:modeldag_extension}) where the linear connection between \texttt{c} and \texttt{m} depends on a non-constant \texttt{beta}. Following the Type Ia Supernovae example, this modification gets close to the dust-to-dust modeling introduced by \cite{popovic2023} ; see also \cite{broutscolnic2021}

Consequently, thanks to the combination of: (1) a simple python dictionary structure, (2) a fully flexible function-based parameter definition approach, and (3) the introduction of the "@" keyword to build any graph structure, the \texttt{modeldag} package enables the user to easily build any model as complex  as desired. In addition, \texttt{modeldag} automatically builds the parameter drawing graph so that the \texttt{model} dictionary can be built and combined in any order. 

\section{\texttt{skysurvey}'s structure and logic}
\label{sec:structure}

\begin{figure*}
\sidecaption
  \includegraphics[width=0.7\linewidth]
  {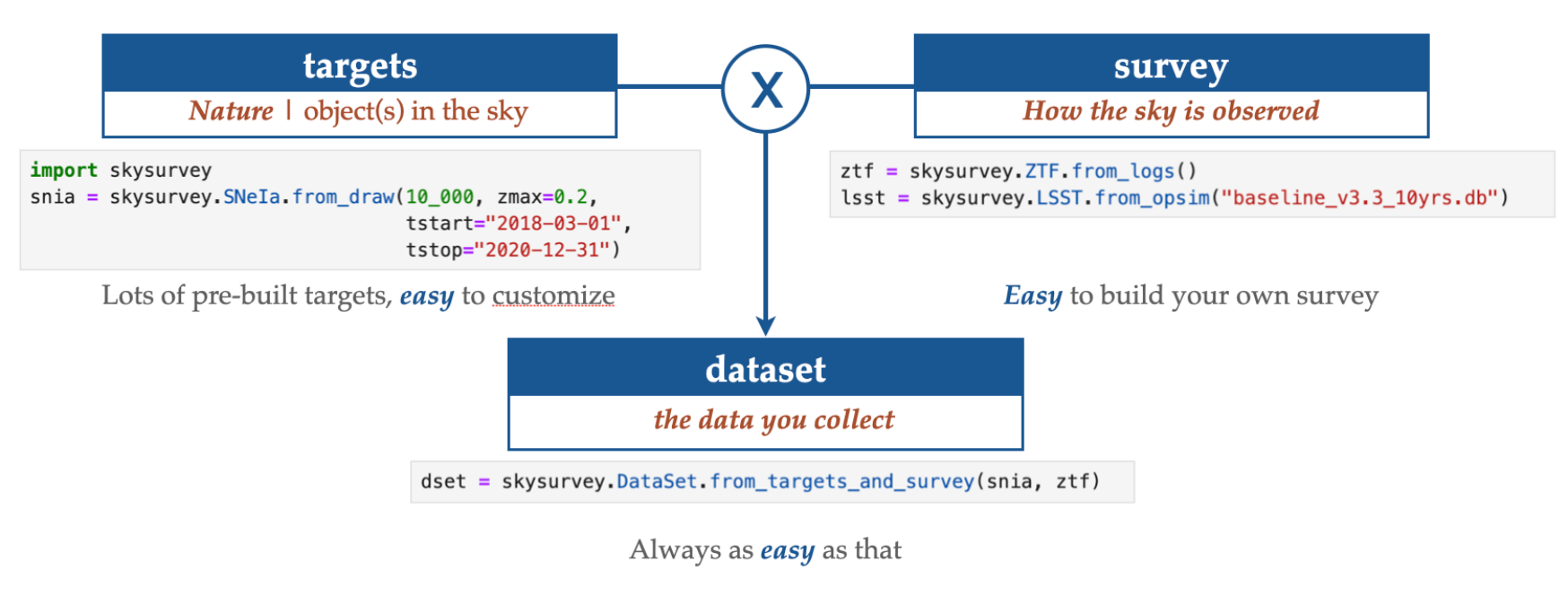}
  \caption{Illustration of the  \texttt{skysurvey} code structure and  associated code. Skysurvey aims at making easy arbitrarily complex realistic simulations of astrophysical targets (e.g. a Supernovae, or any user defined) as observed by surveys (e.g. ZTF, LSST, or any user defined). 
  }
  \centering
  \label{fig:structure}
\end{figure*}

The \texttt{skysurvey} package aims to simulate the transient sky in a forward way. 
Following that logic, the code enables (1) to generate \slobject{Target} objects as given by nature, i.e. noise- and selection-free (see Section~\ref{sec:target}) ; (2) to create \slobject{Survey} objects containing observing information (which filter has been used when to observe where under which conditions ; see Section~\ref{sec:survey}) ; 
 (3) to combine both objects to generate realistic lightcurves. This combination is trivially handled by the \slobject{DataSet} object (see Section~\ref{sec:dataset}). This is illustrated in Fig.~\ref{fig:structure}.

In the following subsection, we detail how these three objects work. The \texttt{skysurvey} package has other  objects that are used internally and will be described when relevant within these subsections.

\section{\texttt{Target}}
\label{sec:target}

A \slobject{skysurvey.Target} represents the astrophysical objects to be simulated. The key attributes of this family of objects are: \texttt{template} and \texttt{model}. 

The \texttt{template} corresponds to the spectro-photometric model from which a lightcurve or a spectrum could be generated. In its released version, \texttt{skysurvey} relies on \texttt{sncosmo} for handling templates \citep{barbary2016}. This is handled through the \slobject{skysurvey.Template} object that manages  interactions between other \texttt{skysurvey} objects and \texttt{sncosmo.Model}s (so \texttt{sncosmo.Source}s ; see \href{https://sncosmo.readthedocs.io/en/stable/}{\texttt{sncosmo}'s documentation}). 

As such, any \texttt{sncosmo.Source} (built-in or user-defined) can be used in \slobject{skysurvey}.

The \texttt{model} attribute is a \texttt{modeldag} object (see Section~\ref{sec:modeldag}) made to simulate target's parameters. These parameters are, at least, those needed by the \texttt{template} and sky coordinates (\textsc{RA} and \textsc{DEC}) needed to match the target with a survey and thereby to simulate realistic observations of the target (see Section~\ref{sec:dataset}).

\subsection{Simulating targets}
\label{sec:target_draw}

Once the \texttt{Target} is loaded, one can use the \texttt{draw()} method to generate a sample. This method  internally calls the \texttt{draw()} method of \texttt{modeldag} (i.e. of the \texttt{model} attribute) to generate a pandas DataFrame where each row corresponds to an individual target and the columns are the parameters specified in the \texttt{model}. This dataframe is stored as a \texttt{Target.data} attribute. 

Up on calling of the \texttt{.draw()} method, the user can alter the model or the model's parameters, such as the redshift range of the  targets and their corresponding time-range (corresponding to \texttt{sncosmo}'s \texttt{t0} source parameters). This functionality fully exploits the  modularity of the dictionary-oriented structure of \texttt{modeldag} at the core of the \texttt{skysurvey} package.

Target redshifts are generated following the \texttt{Target.rate} attribute which could either be constant or a function of redshift or of  any parameter such as the host properties (see detail in Section~\ref{sec:target_rates}). 

Finally, given the redshift and time-range you can either specify the number of target you want (e.g. \texttt{draw(size=10\_000)}) or leave the code estimate the sample size given the volume of the Universe covered by the redshift range, the target rate, the cosmology, and the time of observation. In addition, since \texttt{Survey} could be focusing on narrow region of the sky (e.g. few deep-fields), the user is invited to provide the \texttt{skyarea} (\texttt{skyarea = survey.get\_skyarea()}) as input of the \texttt{draw()} method. This \texttt{shapely} (Multi)Polygon object limits the area of the sky where targets will be drawn (\citealt{gillies2023} ; see Section~\ref{sec:survey}). This is particularly useful when simulating pencil-beam observations like deep-fields ; no need to generate millions of never-observed targets.

\subsection{rates and redshifts}
\label{sec:target_rates}

The rate at which a transient occurs is a crucial input for any simulation, as it specifies how redshifts are drawn. 
The \texttt{skysurvey} package relies on volumetric rates in $\mathrm{target}/ \mathrm{Gpc}^3/\mathrm{yr}$ which could be specified in two ways: (1) as a constant (float) or (2) as a function. The former assumes that the rate is the same regardless of, say, direction or redshift. The latter enables a full flexibility of the rate: it could depend of redshift \citep[e.g. ][]{frohmaier2019}, and/or host properties \citep[e.g.][]{sullivan2006} etc. In any case, the rate function is expected to return the target's volumetric rate in  $\mathrm{target}/ \mathrm{Gpc}^3/\mathrm{yr}$.

To draw redshifts, the code first defines redshift bins (flexible size, but $\Delta z= 10^{-4}$ by default), second computes the Universe's volume ($\mathrm{Gpc}^3$) of this redshift shell given a cosmology, third computes the volumetric target rate at the pivot redshift of the shell. Fourth, the target redshifts are randomly drawn from these redshift bins weighted by the number of targets per bin.

\subsection{Collections of targets}
\label{sec:target_collections}

The \texttt{skysurvey} package enables the user to build \texttt{Target} based on any \texttt{sncosmo} template (e.g. SALT-like, or SN templates) or, even more generally, any user-defined \texttt{sncosmo.Source}. 

However, for most of transients, such as Type II Supernovae, there is no generic template (unlike SALT for SNe Ia) but rather a variety of templates based on well-observed objects. For instance, \texttt{sncosmo} contains 23 templates for Type II Supernovae as defined by \cite{vincenzi2019}.

To simplify the user experience, a \texttt{MultiTemplateTSTransient} object has been implemented in \texttt{skysurvey} that behaves just like a \texttt{TSTransient} but handles the multiple-template issue in the back-end. Based on that, we have pre-built a variety of multiple-template targets, such as Type II Supernovae (see a list in Section~\ref{sec:builtintransients}).

\begin{figure}
  \includegraphics[width=1\linewidth]{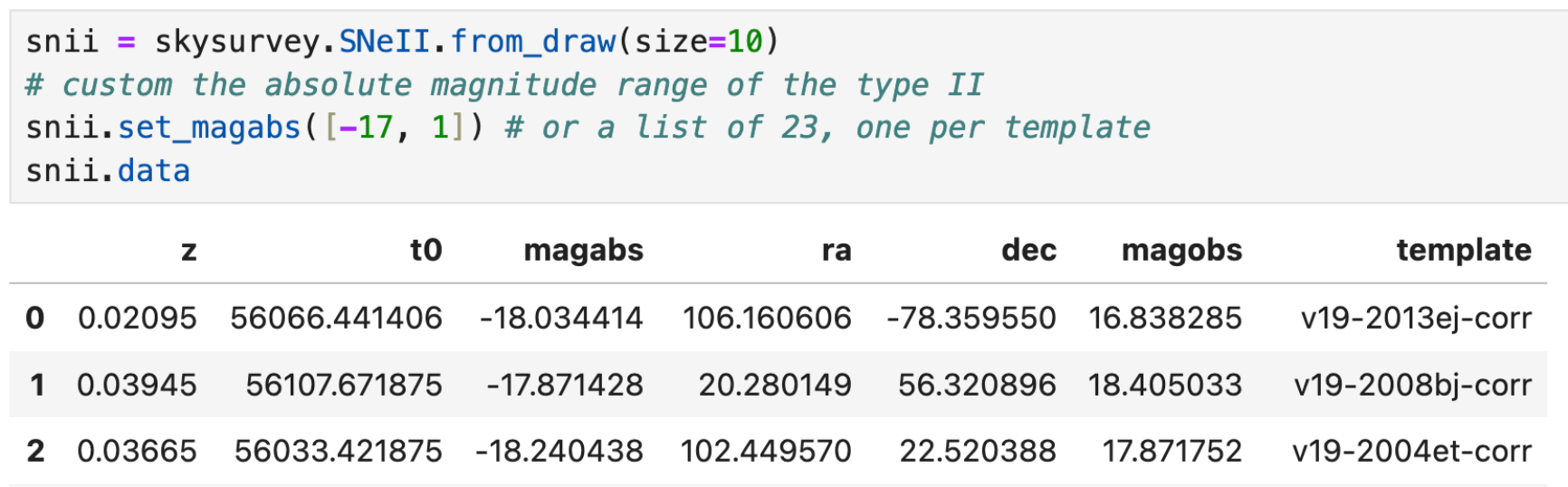}
  \caption{Illustration of how to use multi-template based objects such as the Type II Supernovae \texttt{SNeII}. The actual individual target (dataframe row) template is randomly drawn respecting the rate or model parameters that could be template-dependent.
  }
  \centering
  \label{fig:multitransients}
\end{figure}

As such, the \texttt{skysurvey.SNeII} target object contains, as a \texttt{template}, these 23 \cite{vincenzi2019} templates. 
When the user calls the \texttt{.draw()} method, the individual target's template is drawn alongside all other parameters defined by the object's \texttt{modeldag} (see Section~\ref{sec:target}). 
To draw these templates, the code takes into account that the relative rate of each of these templates may vary. To specify this, the \texttt{rate} parameter of a \texttt{MultiTemplateTSTransient} object can either be a constant, a function for any normal \texttt{Target},  a list of constants, or a list of functions respecting the numpy broadcasting rule (e.g. a list of 1 or 23 constants in that case). In the former case (or if the list has one entry) the templates are equiprobable. This is illustrated in Fig.~\ref{fig:multitransients}.

In addition, the \texttt{skysurvey} package also contain a generic \texttt{TargetCollection} object that can be instantiated by giving a list of any \texttt{Target} (or equivalent such as a \texttt{MultiTemplateTSTransient}). The package is made such that this \texttt{TargetCollection} works anywhere a \texttt{Target} object would ; most notably, such a target collection is accepted as input of  \texttt{DataSet.from\_target\_and\_survey()} (see Section~\ref{sec:dataset}).

\subsection{built-in transients}
\label{sec:builtintransients}

\texttt{skysurvey} comes with a number of built-in \texttt{Target} objects (see Table~\ref{tab:builtin_target}), and each can be customized and modified at will.

In practice, these objects are built on the \texttt{Transient} virtual class that contains all the core functionalities. These built-in \texttt{Target} inherit from this \texttt{Transient} class and pre-define a few attributes, mostly the default \texttt{model}, the \texttt{rate} and the \texttt{template} ; which all can easily be extended or modified by the user (see Section~\ref{sec:modifytarget}).

\begin{table}
\centering
\tiny
\caption{Built-in \texttt{Target} object of the \texttt{skysurvey} package}
\label{tab:builtin_target}
\begin{tabular}{l c c}
\hline\\[-0.8em]
name  & kind & template(s) \\[0.30em]
\hline\\[-0.8em]
\hline\\[-0.5em]
\texttt{SNeIa} & \texttt{Transient} & SALT-like \\[0.15em]
\texttt{TSTransient} & \texttt{Transient} & any \texttt{sncosmo.TimeSerie} \\[0.15em]
\texttt{SNeIb} & \texttt{MultiTemplateTSTransient} & 12 templates*  \\[0.15em]
\texttt{SNeIc} & \texttt{MultiTemplateTSTransient} & 7 templates*  \\[0.15em]
\texttt{SNeIcBL} & \texttt{MultiTemplateTSTransient} & 6 templates*  \\[0.15em]
\texttt{SNeII} & \texttt{MultiTemplateTSTransient} & 23 templates*  \\[0.15em]
\texttt{SNeIIb} & \texttt{MultiTemplateTSTransient} & 12 templates*  \\[0.15em]
\texttt{SNeIIn} & \texttt{MultiTemplateTSTransient} & 6 templates * \\[0.15em]
\texttt{Kilonovae} & \texttt{Transient} & POSSIS**  \\[0.15em]
\hline
\end{tabular}
\tablefoot{*: \texttt{sncosmo.TimeSerie} defined by \cite{vincenzi2019} \\ **: \texttt{sncosmo.AngularTimeSeriesSource}, model from \citep{bulla2019}}
\end{table}

\subsection{Modifying a target}
\label{sec:modifytarget}

The \texttt{rate} and \texttt{model} \texttt{Target} attributes can be changed to modify the number of targets expected as a function of redshift, or how the target parameters are drawn. These changes can be either when instantiating the target (options of the \texttt{.from\_draw()} class method) or on a loaded target using the associated \texttt{.set\_rate}, \texttt{.set\_model} or \texttt{.update\_model[\_parameters]}. The \texttt{update\_model} method fully exploit the \texttt{modeldag} dictionary structure as explained in Section~\ref{sec:modeldag}. With this method you only change how parameters are drawn and/or connected together, or even add new parameters.  As explained in Section~\ref{sec:modeldag}, a new graph will automatically be built once the model is updated. The \texttt{.update\_model\_parameters} method directly affects the \texttt{kwargs} entries on the model to change the arguments of the function used to draw a variable, but not the functional form itself.

\begin{figure}
  \includegraphics[width=1\linewidth]{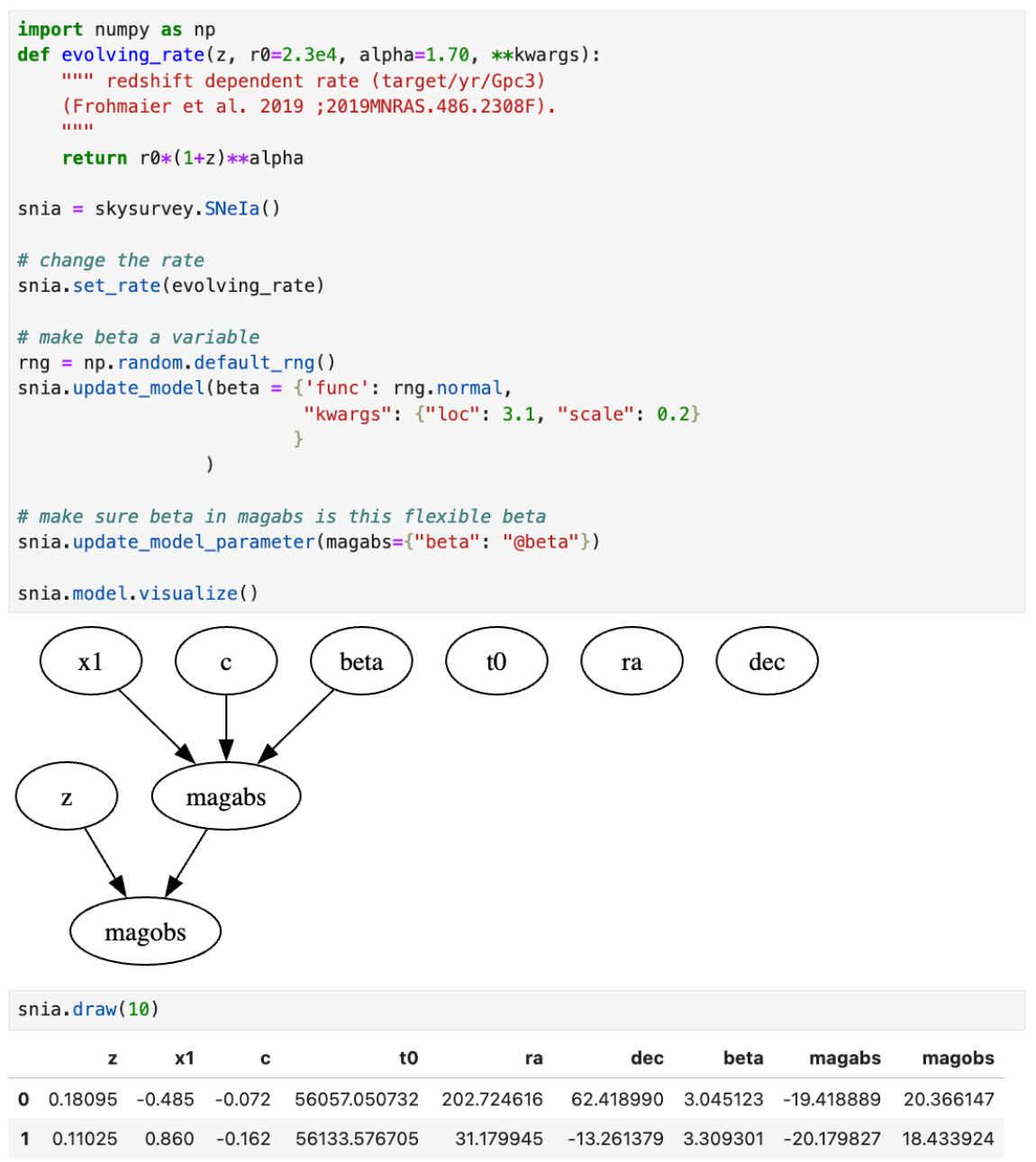}
  \caption{Illustration of how to change a target ; here, specify a redshift dependent \texttt{rate} and make SN Ia standardization parameter \texttt{beta} a variable drawn from a normal distribution.
  }
  \centering
  \label{fig:changingtarget}
\end{figure}

This is illustrated in Fig.~\ref{fig:changingtarget}. 
In that example, we change the default \texttt{SNeIa} rate (a constant volumetric rate of $2.35\times10^4$ $\mathrm{target}/ \mathrm{Gpc}^3/\mathrm{yr}$, \citealt{perley2020}) to a redshift-evolving rate as modeled by \cite{frohmaier2019}. In addition, we make the \texttt{beta} standardization parameter of the SNe Ia Tripp's relation a normally distributed variable instead of a usual constant. 

As illustrated in this Section, the \texttt{skysurvey} package, thanks to the \texttt{modeldag} structure, has been designed to make easy complex / realistic simulations. The user is only limited by its imagination and ability to define how the target's parameters should be drawn and connected together, i.e. setting the \texttt{model} and/or the \texttt{rate}'s function. The rest is automatically handled by \texttt{skysurvey}.

\subsection{Adding effects}
\label{sec:effect}

\texttt{skysurvey} provides a flexible mechanism for adding various physical effects to the simulation. This is handled by the \slobject{Effect} class, which can modify both the intrinsic properties of the transients and/or the way their lightcurves or spectra are generated.

In practice, a \texttt{skysurvey.Effect} has two core attributes: 1. \texttt{.effect}, an \texttt{sncosmo.PropagationEffect}, which alters how transient fluxes (mag) are generated, be it when getting a lightcurve or a spectrum ; 2. \texttt{.model}, a \texttt{modeldag} dictionary that will be used to update the target's model to generate the \texttt{.effect} parameters if any. 

An effect can be added to any target either during instantiation, using the \texttt{effect=} option of the \texttt{.from\_draw()} class method, or directly using the \texttt{.set\_effect()} method. A target can handle multiple effects simultaneously.

\subsubsection{Milky Way}
\label{sec:effect_milkyway}

A good example of this is the handling of Milky Way dust extinction. \texttt{skysurvey} provides a pre-defined \slobject{Effect} easily accessible by calling \texttt{mw = skysurvey.Effect.from\_name("mw")}. 

To handle Milky Way extinction, an observer-frame dust-like \texttt{sncosmo.PropagationEffect} is generated \citep[][by default]{ccm1989}, which has two parameters, "mwebv" and "rv" (3.1 by default). 
The "mwebv" parameter refers to the intensity of the Milky Way dust along the line of sight and hence depends on the target's coordinates. 
Consequently, \texttt{mw.model} has a \texttt{modeldag} dictionary with \texttt{func} a custom made \texttt{get\_mwebv(ra, dec, which='planck')} function (relying on the \texttt{dustmaps}\footnote{\url{https://dustmaps.readthedocs.io/en/latest/}} package; \citealt{green2018}) and \texttt{kwargs: dict(ra='@ra', dec='@dec')} to pass the target's coordinate information to the \texttt{get\_mwebv()} function.
When the \texttt{mw} is added to the target, \texttt{mw.model} is used to update the \texttt{target.model}, which enables to generate the 'mwebv' parameter upon the \texttt{.draw()} call, and the \texttt{sncosmo.PropagationEffect} is passed to the template to handle the effect on flux generation.

\begin{figure}
  \includegraphics[width=1\linewidth]{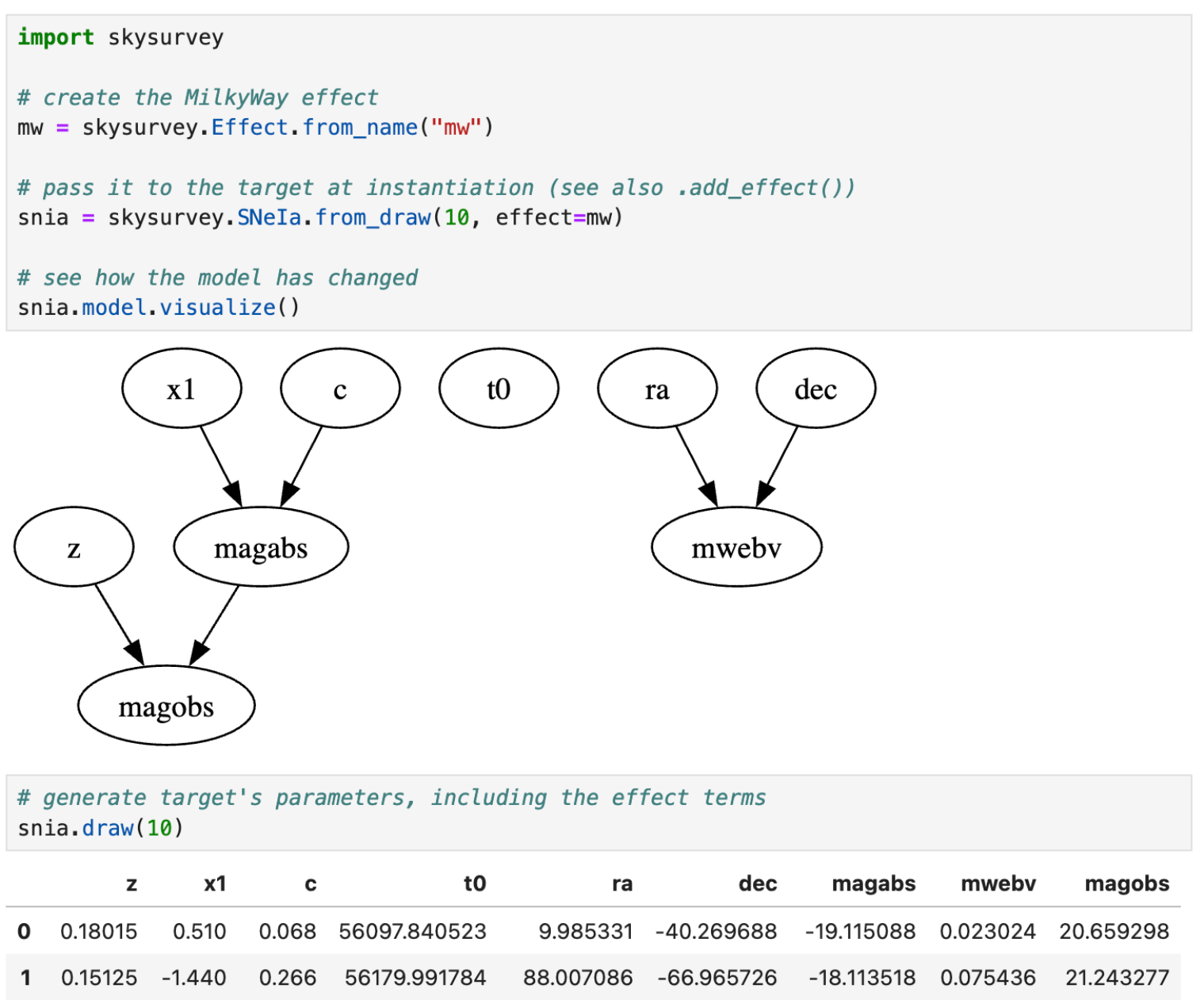}
  \caption{Illustration of how to add an effect to a target ; here, the Milky Way extinction, whose intensity parameter (mwebv) depends on the target's coordinate. 
  }
  \centering
  \label{fig:milkyway_effect}
\end{figure}

This is illustrated in Fig.~\ref{fig:milkyway_effect} which demonstrates the simplicity for the user to account for Milky Way extinction. 

For SNe Ia, another useful effect is that associated to color scatter. Those kind of effect have no model parameters, just a propagation effect. 
By default, two scatter effects have been implemented, 'C11' \citep{chotard2011} and 'G10' \citep{guy2010}. These can be loaded using the \texttt{cscatter = skysurvey.Effect.from\_name('scatter', 'c11')}.

Creating new customed effects thus simply requires the user to define a \texttt{sncosmo.PropagationEffect} and the \texttt{modeldag} model that draws the propagation effect parameters, potentially depending on other parameters, just like for the Milky Way effect.

\section{\texttt{Survey}}
\label{sec:survey}

The \slobject{Survey} objects in \texttt{skysurvey} represent a time-domain survey. A survey is defined by a spatial footprint of the camera in the sky, and observing logs (or corresponding simulation), specifying what part of the sky has been observed when and under what conditions. Under the hood, \texttt{Survey} objects are made both highly modular while remaining extremely fast in matching observing logs with any sky location (hence any target) by a combination of \texttt{shapely}\footnote{\url{https://shapely.readthedocs.io/en/stable/manual.html}} and \texttt{geopandas}\footnote{\url{https://geopandas.org}} libraries and, for some, by using the healpix decomposition \footnote{\url{https://healpy.readthedocs.io/en/latest/}}. 

There are two kind of surveys in \texttt{skysurvey}: the generic \texttt{skysurvey.Survey}, and the more specific \texttt{skysurvey.GridSurvey} for surveys with fixed pointing grids, like, e.g. ZTF or deepfield surveys like DES, to speed up the computations. 

In both cases, the \texttt{(Grid)Survey.footprint} is a shapely (Multi)Polygon object in degrees. The MultiPolygon case is particularly useful for multi-detector camera, such as most of the modern ones. There, each polygon of MultiPolygon corresponds to a \texttt{ccdid} (customable name), that has matching entries in the observing logs. As such, the observing condition are per ccd (or read-out channel, etc. as defined in the log) which could vary even within a single exposure. 

The observing logs is stored as a pandas.DataFrame (\texttt{survey.data}) and must contain, at least: the point coordinates \texttt{ra}, \texttt{dec} (or the \texttt{fieldid} for GridSurveys) ; \texttt{mjd}, the observing time in modified Julian dates ; \texttt{band}, the filter used to observe the sky ; \texttt{skynoise}, the noise (in ADU) of that log entry, \texttt{gain}, to convert ADU to electron when computing the Poisson noise ; \texttt{zp}, the zeropoint of the entry. 

The coordinate's information, together with the survey footprint, is used to match any target with the survey logs. This is made using the \texttt{survey.radec\_to\_fieldid()} method, which uses the spatial join \texttt{geopandas} method for \texttt{GridSurvey}s (e.g. ZTF) or healpix pixelisation for generic \texttt{Survey}s (e.g. LSST). In both cases, matching 100k targets with a full sky survey takes only a fraction of seconds. This speed is one of the key aspects of \texttt{skysurvey}. 

The online documentation shows several examples on how to create a customed \texttt{Survey} or a \texttt{GridSurvey}. In Section~\ref{sec:builtinsurvey} we illustrate each kind with the pre-built \texttt{skysurvey.ZTF} and \texttt{skysurvey.LSST} surveys.

\subsection{Built-in Surveys}
\label{sec:builtinsurvey}

While \texttt{skysurvey} makes it easy to build any kind of surveys, a few major surveys have been built-in and more will be added in the future. We invite the user to see online documentation for an up-to-date list. In this paper, we present ZTF and LSST's implementation as they cover both survey kinds: a GridSurvey (ZTF) and a generic Survey (LSST). 

\subsubsection{ZTF: a \texttt{GridSurvey}}
\label{sec:ztf}

\begin{figure}
  \includegraphics[width=1\linewidth]{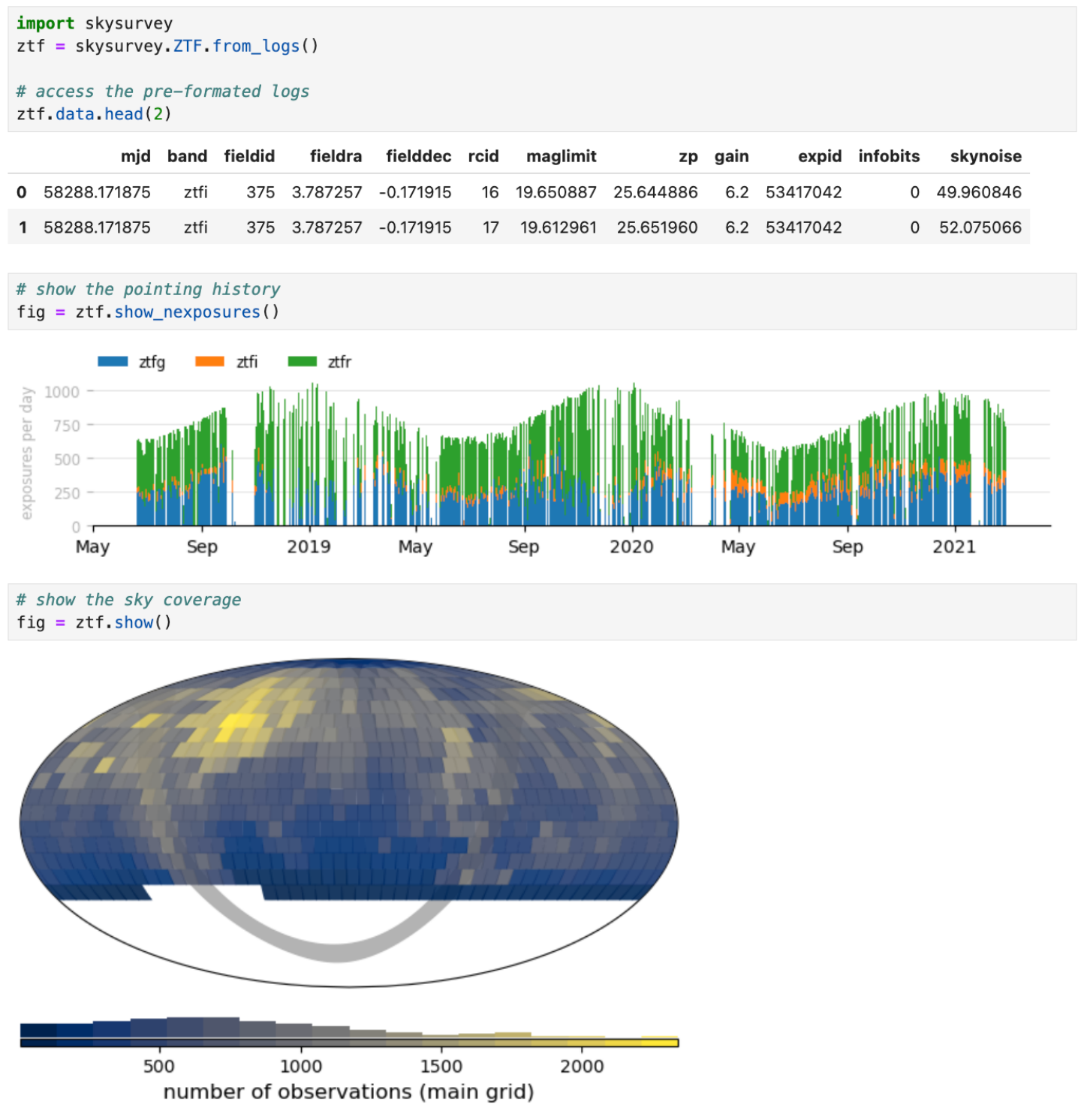}
  \caption{Illustration of how to load the \texttt{ZTF} survey and basic plotting functionalities. The grid system of ZTF is clearly visible in the bottom panel that shows the number of visit for the main grid. The grey band illustrates the Milky Way location. The northern ZTF survey does not cover $\mathrm{dec}<-30$.}
  \centering
  \label{fig:ztf}
\end{figure}

The ZTF survey acquires observations following two pre-defined pointing grids \citep[see e.g.][]{bellm2019,masci2019}. The main grid has \texttt{fieldid<1000} while \texttt{fieldids>1000} are those of the secondary grid. This griding corresponds to the camera footprint in the sky as illustrated, for the main grid, in Fig.~\ref{fig:ztf}. The secondary grid is offset by a quarter of camera footprint and is used to increase the full sky coverage of the survey and for calibration purposes. In practice, this secondary grid is only acquired 10\% of the time.

The central (\texttt{ra}, \texttt{dec}) coordinates for each \texttt{fieldid} is publicly available and easily accessible through the \texttt{ztffield}\footnote{\url{https://ztffields.readthedocs.io}} package. 
In \texttt{skysurvey}, we use this package to build the \texttt{GridSurvey.fields} attribute that contains the \texttt{shapely.polygon} corresponding to the \texttt{footprint} centered in all \texttt{fieldid} coordinates. As such, logs for \texttt{GridSurvey} like ZTF only need to list which field has been observed, without having to specify the pointing coordinates (ignored if given).

The \texttt{ZTF} survey has been implemented in \texttt{skysurvey} such that the user does not need to pay attention to all these details. We have also implemented  \texttt{.from\_logs()} class method to instantiate ZTF given the latest ZTF survey log, i.e., these from \cite{rigault2025} when writing this paper. This is illustrated in Fig.~\ref{fig:ztf}.

The \texttt{ZTF} \texttt{GridSurvey} has a multi-polygon \texttt{footprint} based on its 64 quadrants identified by an \texttt{rcid} \citep[see log description in][]{rigault2025}.

\subsubsection{LSST: a \texttt{Survey}}
\label{sec:lsst}

The LSST survey does not follow a pre-defined pointing grid but points anywhere in the sky and the observing "logs" (in practice \texttt{opsim} simulations when this paper is being written) hence contain these (\texttt{ra}, \texttt{dec}) coordinates, corresponding to the pointing location of the center of the camera. Then, assuming the LSST footprint (currently a single polygon, hence ignoring single ccd-variations) one can compute which part of the sky has effectively been observed as a function of time. We project information into a healpix sky pixelisation using the \texttt{healpy} package. By default, the \texttt{skysurvey} packages is set to use $\texttt{n-side}=200$ healpix-pixelisation, corresponding to 480k sky pixels of 0.086 deg$^2$ each ; this is tunable up load instantiation of the \texttt{Survey}.

\begin{figure}
  \includegraphics[width=1\linewidth]{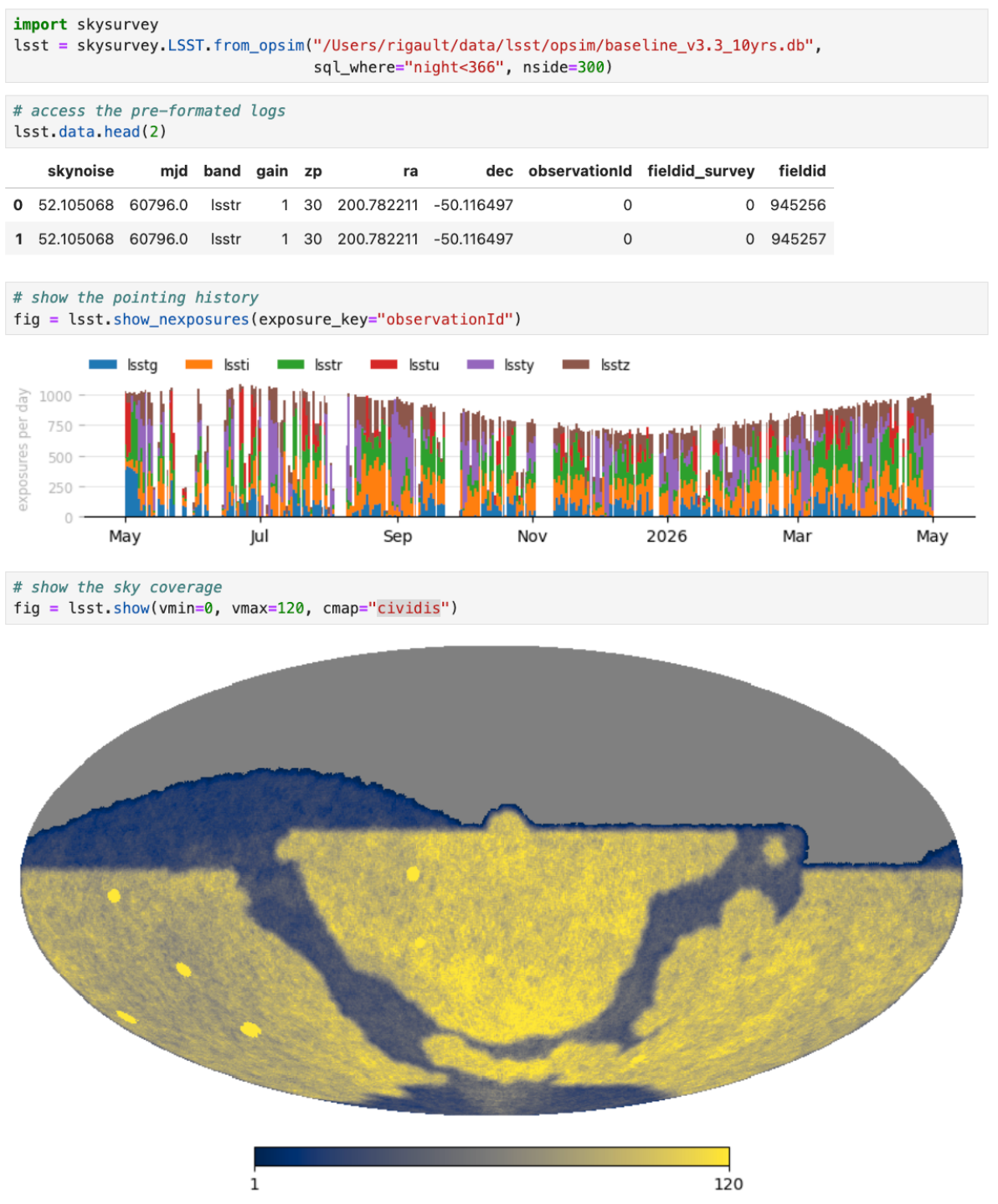}
  \caption{Illustration of how to load the \texttt{LSST} survey, presenting how to select opsim rows while loading the data (here only the first year of data) and changing the healpix sampling (\texttt{nside}). Functionalities of survey and gridsurvey (see Fig.~\ref{fig:ztf}) are self-consistent, e.g., the plotting methods as shown here.}
  \centering
  \label{fig:lsst}
\end{figure}

We illustrate Fig.~\ref{fig:lsst} how a generic survey like LSST can be loaded and visualized. In this example, we show that one can select only the first year of data while changing the healpix pixelisation ($\mathrm{nside}=300$, so 0.038 deg$^2$ pixels). 

Figures~\ref{fig:ztf} and~\ref{fig:lsst} illustrate that both kind of surveys have the same functionalities and, in practice, the user does not need to know which kind of survey the code uses; \texttt{skysurvey} handles them self-consistently. Notably, \texttt{.data} contains the observing logs, \texttt{.radec\_to\_fieldid} to match target location with entry logs (\texttt{fieldid} corresponding to the healpix index for \texttt{Survey}), \texttt{.get\_skyarea} that returns the total survey footprint in the sky in form of a \texttt{shapely} geometry (see Section~\ref{sec:dataset}), \texttt{.get\_timerange()} to get the starting and ending dates of the survey (see Section~\ref{sec:dataset}), and, finally, plotting functions as illustrated in these Figures~\ref{fig:ztf} and~\ref{fig:lsst}.

\section{\texttt{DataSet}}
\label{sec:dataset}

The \slobject{DataSet} object is the final product of a \texttt{skysurvey} simulation. 
This object simulates the lightcurves by matching an input \texttt{target} (or list of) with the survey logs. The lightcurves generation is detailed in Section~\ref{sec:generatinglc} and  key \slobject{DataSet} methods are presented in section~\ref{sec:datasetmethods}. We show in section~\ref{sec:example} how the \slobject{DataSet} object can be used to accurately simulate a survey like the ZTF Type Ia Supernovae dataset \citep{rigault2025}.

\subsection{Generating lightcurves}
\label{sec:generatinglc}

The lightcurve generation is made at instantiation of the \texttt{DataSet} using the \texttt{.from\_targets\_and\_survey()} class method. This method, the backbone of the \texttt{DataSet} object, works as follows.

First, targets are associated with the survey's logs using the \texttt{.radec\_to\_field()} survey method, which takes a fraction of second to provide the \texttt{fieldid} (i.e., the healpix pixel id for generic survey) associated to each target. In case of a \texttt{GridSurvey}, a target can have several \texttt{fieldid}s as fields can overlap, and this \texttt{fieldid} could be a multi-layer index, especially if the survey \texttt{footprint} is decomposed by detectors or amplifier (see Section~\ref{sec:survey}). At this stage, we have a way to select the entry logs corresponding to any given target and we know which target has no entry (e.g. simulated outside the survey sky-footprint).

Second, we loop over each target and we build its \texttt{sncosmo.Model} including its template parameter alongside with any additional effect (e.g. the Milky Way dust extinction or color scatter, see Section~\ref{sec:effect}). This is made using the \texttt{target.get\_target\_template()} method and takes about 1$ms$ per target.

Third, once the target \texttt{sncosmo.Model} is loaded, we simulate target flux using the \texttt{model.bandflux} method given the "band", "mjd" and "zp" survey logs entry, and we compute the flux error given the "skynoise", this simulated "flux" and the camera "gain". Optionally, this computation is only made in the $[-50, +200]$ d rest-frame phase range around the model's $t_0$ (explosion date or peak-date for SALT-like models). This data is stored as a pandas dataframe and that of each target are concatenated into a multi-index \texttt{pandas.DataFrame} with the first index for the target and the second for the survey log entry. This multi-index dataframe then become the \texttt{.data} attribute of the \slobject{DataSet}. 

Altogether, on a laptop, this whole operation, from matching each target with the survey logs to the storage of the massive multi-index dataframe, takes about 5min for 100k targets (~350 targets processed per second) for a $3\pi$ sky survey like \texttt{ZTF} or \texttt{LSST}. 
The code supports dask parallel computing to distribute this computation on a cluster (especially interesting for $\mathcal{O}(10^{6+})$ targets simulations) and work is ongoing to replace \texttt{sncosmo} for the flux generation to enable vectorization; 
10x speed up is expected with a aim of simulating 100k target in less than 1min.

\subsection{Using a \texttt{DataSet}.}
\label{sec:datasetmethods}

The \texttt{skysurvey.DataSet} object has three main attributes: \texttt{.targets}, a pointer to the input (multi-)target object; \texttt{.survey}, a pointer to the input survey; and \texttt{.data}, the multi-index dataframe containing the simulated lightcurves (see Section~\ref{sec:generatinglc}). Based on these, and mostly on \texttt{.data}, the \texttt{DataSet} has top level functionalities to ease the user experience. 

\begin{figure}
  \includegraphics[width=1\linewidth]{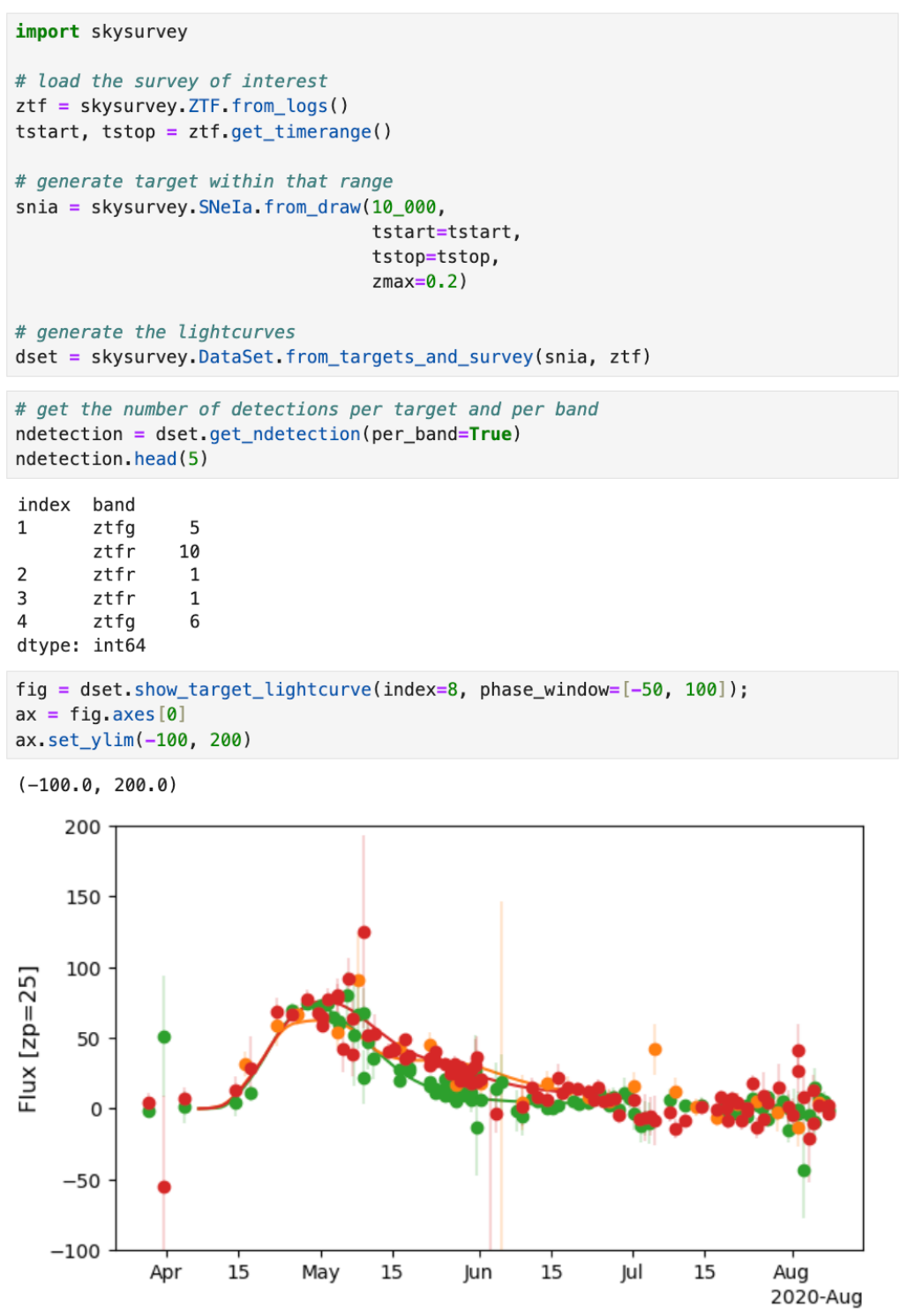}
  \caption{Illustration of an end-to-end \texttt{skysurvey} simulation, from top to bottom: building a survey (\texttt{ZTF}), generating targets (\texttt{SNeIa}) in the time range of the survey and building the \texttt{DataSet} object that simulate the lightcurve. Then, in the next cells, illustrations of methods to access lightcurves statistics and visualize a simulation. This whole sequence takes $\sim15\ \mathrm{s}$ to run on a laptop (see text).}
  \centering
  \label{fig:dataset}
\end{figure}

First, the \texttt{.get\_target\_lightcurve()} method enables the user to directly extract the lightcurves of any observed target while applying a variety of cuts (.e.g as a function of flux signal to noise or time) or while adding extra information (e.g. the phase). Second, the \texttt{.get\_ndetection()} method enables the user to have access to top level statistics associated with each lightcurve. As illustrated in fig.~\ref{fig:dataset}, this method returns the number of detections ($\mathrm{flux}/\mathrm{flux_{err}} > n$ ; with $n=5$ by default) per target (and per band, optionally), combining (optionally) same day detections.
As illustrated in the example analyses (section~\ref{sec:example}), where we reproduce the ZTF SN Ia DR2 dataset \citep{rigault2025}, this method is particularly useful when simulating the selection function of a survey as one usually requests a target to have at least $n$ detections, including at least $m$ pre-max and $p$ post-max and, altogether at least $b$ bands.

\section{Example: reproduce the ZTF SN Ia DR2 dataset}
\label{sec:example}

To illustrate an application case of the  \texttt{skysurvey} package, we present in that section how to reproduce ZTF SN Ia DR2 Type Ia Supernovae sample. ZTF SN Ia DR2 data are publicly available \citep{rigault2025}, which includes observations logs of ZTF science images from mid-2018 to end-2020. This example, detailed in the following section, is also available on the online \texttt{skysurvey} documentation.

\subsection{Altering survey data}
\label{sec:example_changelogs}

According to \cite{rigault2025b} and \cite{amenouche2025}, the magnitude limits associated with ZTF science images do not fully reflect the depth of difference images from which ZTF SN Ia DR2 lightcurves are derived. They do not account for errors in the reference image and thus under-estimate the sky-noise. Following their recommendations, the \texttt{skynoise} derived from science image  magnitude limits should be increased by 23\%, 17\% and 20\% for $g$, $r$, and $i$ ZTF bands, respectively. 

\begin{figure}
  \includegraphics[width=1\linewidth]{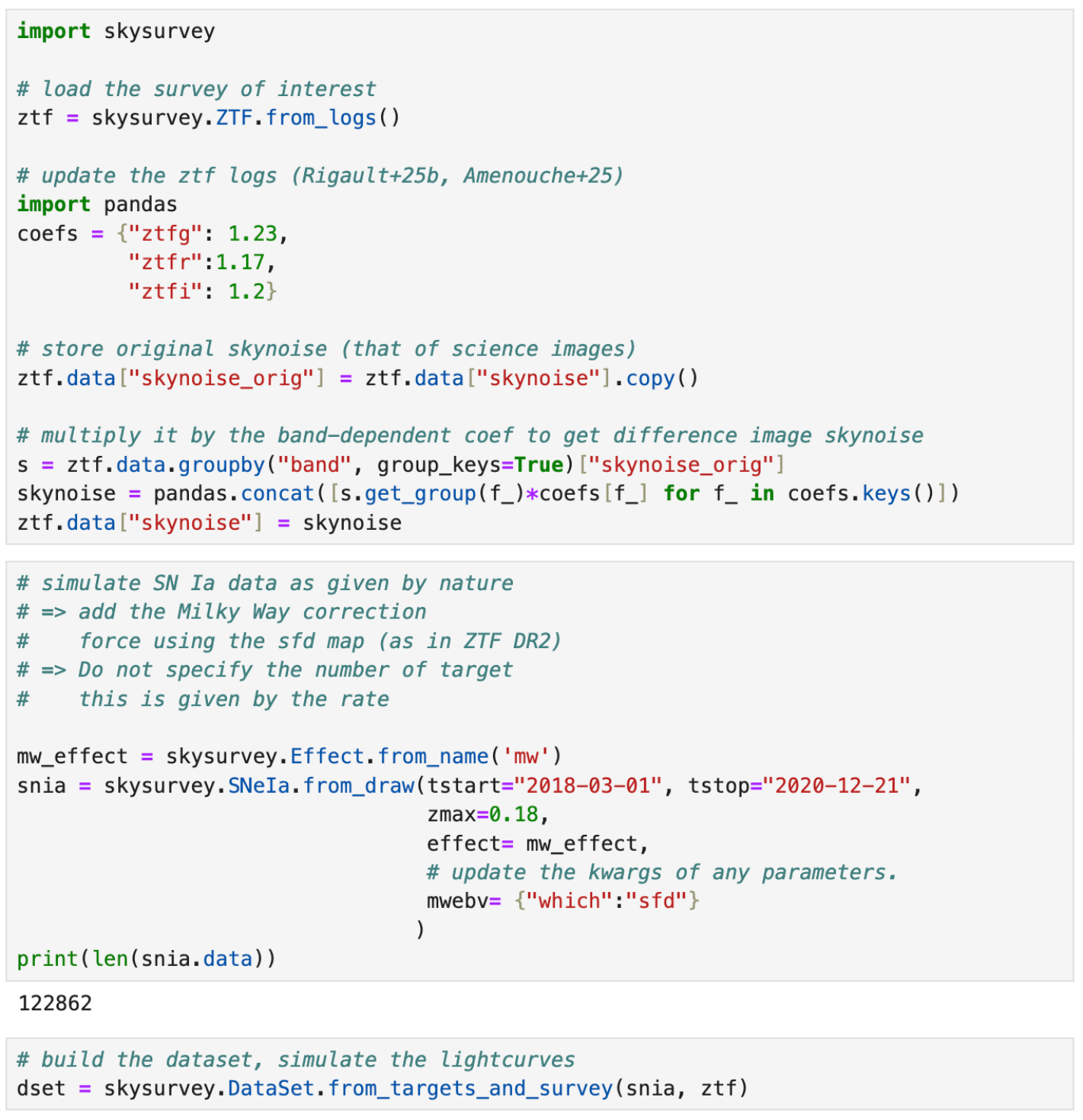}
  \caption{Illustration of how to simulate the ZTF SN Ia DR2 dataset, changing the logs (top panel) and default target (mid panel) to respect prescriptions from the literature (see text). In any case, the dataset generation, which create the actual simulated lightcurve remains unaffected (bottom panel). In that example, the number of target is unspecified and consequently derived from the \texttt{target.rate}.}
  \centering
  \label{fig:simudr2_step1}
\end{figure}

We illustrate in Fig.~\ref{fig:simudr2_step1} (top panel) how this is done with \texttt{skysurvey}. Since the observing logs are stored as \texttt{pandas.DataFrame}, one can use standard \texttt{pandas} procedures to alter the \texttt{survey.data}, in that example, we group the data per band, and we apply to the \texttt{skynoise} entry of each group the corresponding multiplicative coefficient. As such, \texttt{ztf.data["skynoise"]} is now the proper skynoise to consider for the simulation.

\subsection{Loading SNe Ia with Milky Way effects}
\label{sec:example_changesnia}

Once the survey is loaded and ready to be used, we can simulate targets of interest. In \texttt{skysurvey} these targets represent astrophysical events as given by nature. In our example, to reproduce the ZTF SN Ia DR2, we shall consider all SNe Ia given by nature between March 2018 and end-2020 up to a redshift of 0.18 while accounting for dust absorption by the Milky Way. To reproduce this  dataset, we shall  simulate the Milky Way effect using the \cite{sfd1998} dust-map in place of the default Planck one \citep{planck2016} as done in ZTF.

We illustrate in Fig.~\ref{fig:simudr2_step1} how this is done. We explicitly specify the start and end date between which targets should be simulated (i.e., the $t_0$ parameter) and we set the Milky Way dust extinction effect at instantiation of the target. The \texttt{mwebv} parameter hence becomes one of the \texttt{model} parameters and we alter its \texttt{kwargs} using the \texttt{**kwargs} options of \texttt{.from\_draw()} as one can do for any model parameters. 

Notice that the number of targets to be simulated is not explicitly specified. This number (\texttt{size=}) is actually derived from the \texttt{rate} parameter given the volume of the Universe inferred from the input redshift range ($z<0.18$, here), the cosmology (\citealt{planck2020} by default, see options), and the time range of the simulation (March-2018 to end-2020). If the cosmology were to be updated (settable attribute of any \texttt{Target} object), it is self-consistenty passed to the function of any model parameter that requires it (i.e., that have \texttt{cosmo=} or \texttt{cosmology=} in their option). In that example, the Universe should have generated nearly 123k SNe Ia assuming the default rate of $2.35\,10^{4}\ \mathrm{target/Gpc^3/yr}$ from \cite{perley2020}.

\subsection{Generating ZTF SN Ia DR2 lightcurves}
\label{sec:example_dset}

Given any survey and target (or list of) the lightcurve generation is handled by the \texttt{DataSet.from\_targets\_and\_survey} class method. It works as illustrated in Fig.~\ref{fig:simudr2_step1}. On a laptop, this steps takes less than 10 min to simulate 123k SN Ia on the $3\pi$ ZTF survey. As detailed in Section~\ref{sec:dataset} ongoing work is in progress to speed this process, however it can already be parallelized on a cluster using the \texttt{dask} package (see the \texttt{client=} options and online documentation).

\subsection{Compute selection effects}
\label{sec:example_selection}

\begin{figure}
  \includegraphics[width=1\linewidth]{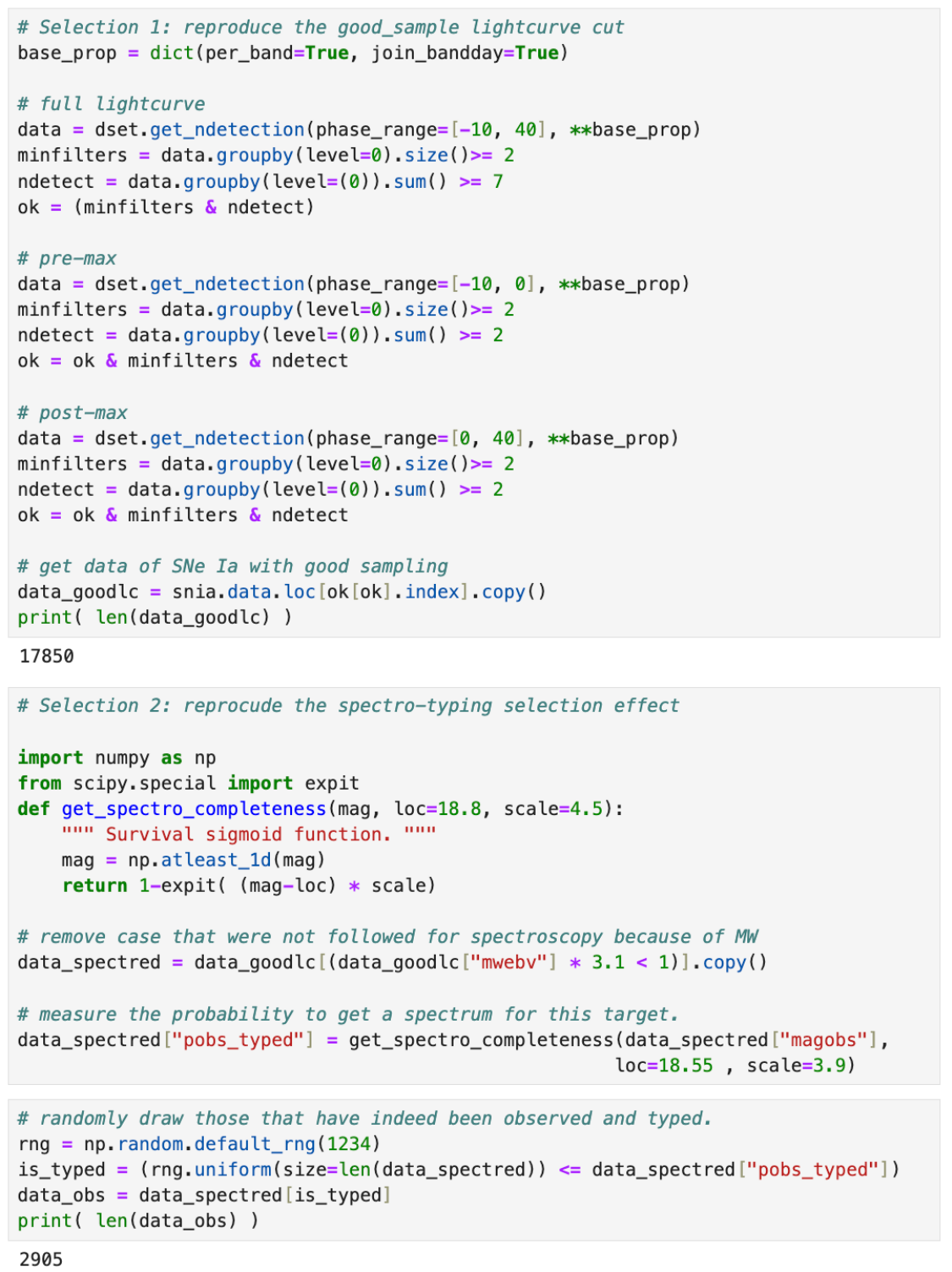}
  \caption{Illustration of how to mimic lightcurve quality cut selection functions (top panel, here that of the ZTF SN Ia DR2 sample from \citealt{rigault2025}) and how to add any kind of selection (here spectroscopic efficiency) to reproduce the expected full well sampled spectroscopically classify SN Ia dataset.}
  \centering
  \label{fig:simudr2_step2}
\end{figure}

\cite{rigault2025} specify that is considered as well sampled target, any SNe Ia that have, in the $[-10, +40]$ d rest-frame phase range: at least 7 5$\sigma$ detections, two of which pre-max, two of which post-max, with at least 2 bands. Same-day detections are ignored to compute this statistics. As illustrated in Fig.~\ref{fig:simudr2_step2}, this cut is easy to do thanks to the \texttt{dataset.get\_ndetection()} method. At this stage, only 18k/123k targets pass the cuts, while 88k/123k are within the ZTF survey footprint (the remaining ones have $\mathrm{dec}<-30$).

On these, we should now apply the spectroscopic typing  selection effect. ZTF did not trigger for spectroscopic follow-up to classify sources with high Milky Way extinction ($\mathrm{A_v}\geq1$, \citealt{perley2020}). Second, the classification selection effects can easily be reproduced by a sigmoid function applied on the peak magnitude \citep{rigault2025}. In the present illustration, we update their sigmoid parameters to $\mathrm{loc}=18.6$ and $\mathrm{scale}=3.9$, in place of 18.8 and 4.5, respectively.

\begin{figure}
  \includegraphics[width=1\linewidth]{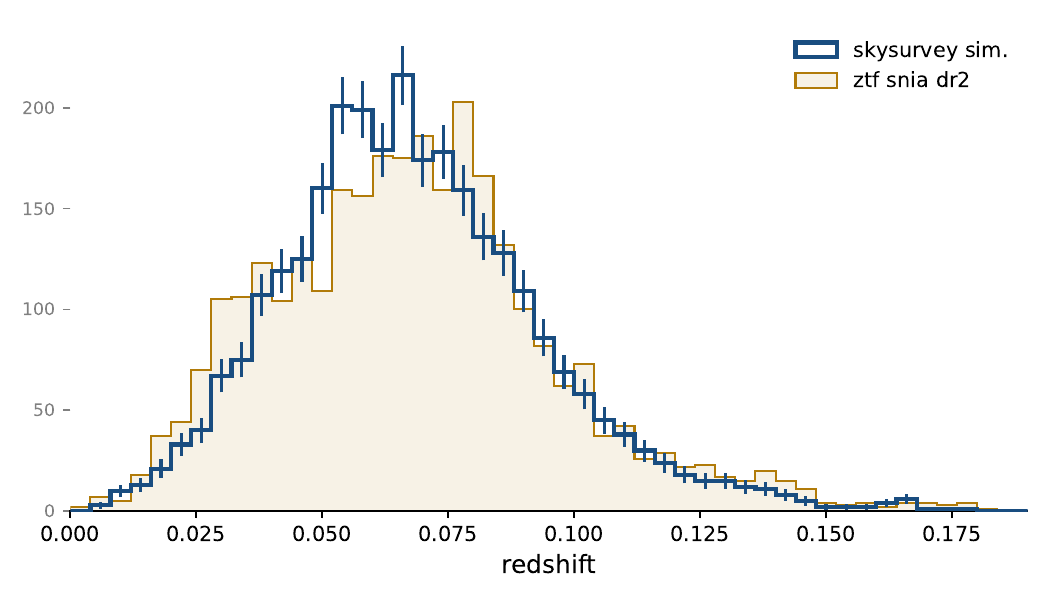}
  \caption{Redshift distribution of the ZTF SN Ia DR2 sample \citep[orange,][]{rigault2025} compared to that simulated using the \texttt{skysurvey} package made without assuming the initial number of target, but directly deriving it from the volumetric rate. Such an analysis illustrates the ability to the \texttt{skysurvey} package to easily reproduce realistic data.
  }
  \centering
  \label{fig:redshift_dr2_skysurvey}
\end{figure}

We finally randomly draw targets depending on their probability to have a spectroscopic classification. This leads to about 2900 targets. In comparison, the ZTF SN Ia DR2 dataset has 2960 well sampled targets \citep[see Table 1 of][]{rigault2025}. We show Fig.~\ref{fig:redshift_dr2_skysurvey} the redshift distribution of the simulated targets overlapped with that actually collected as part of the ZTF SN Ia DR2 release. Both distributions match, proving the accuracy of both the simulation package and the volumetric rate from \cite{perley2020}. 

The apparent excess of ZTF SN Ia DR2 data near $z=0.03$ and the gap by $z=0.05$ are studied in details by \cite{gilles2026}, who show they correspond to local structure in the Universe. 
Such structures, and more generally the Universe large scale structure of the can easily be taken into account \texttt{skysurvey}. To do so, one simply has to build a function, included in the \texttt{Target.model}, and that draws simultaneously $\texttt{ra}$, $\texttt{dec}$ and $\texttt{redshift}$ respecting these structures. This can easily be done by sampling positions from dark matter halo catalogs based on N-body simulations or galaxy catalogs \citep{ishiyama2009, jasche2019, gilles2026}.

\section{Conclusion and Discussion}
\label{sec:conclusion}

This paper presents the \texttt{skysurvey} package. A pure python library that enables the user to easily simulate complex astrophysical populations as they would be observed by a survey. This package has been designed to be fast, intuitive, and highly modular in its ability to build any astrophysical target that could have any parameter modeling structure. 
This later aspect is made possible thanks to \texttt{modeldag}, a standalone package developed in parallel to \texttt{skysurvey} and also presented in this article. 

The \texttt{modeldag} python package provides a way to create a direct acyclic diagram (DAG) between parameters using a trivial python dictionary structure that only has three leafs: \texttt{func} that defines how the parameters should be drawn, \texttt{kwargs} that provides the function arguments, \texttt{as} specifies the name of the returned parameter(s) -- which is particularly useful when returning multiple parameter simultaneously (e.g. coordinates). The trick of \texttt{modeldag} is the introduction of the "@" keyword in \texttt{kwargs}, such that  "\texttt{x='@a'}" is parsed by the code as "use the output of the \texttt{a} parameter draw as \texttt{x} input to draw this other parameter". This way, \texttt{modeldag} automatically creates DAG between parameters while offering an entire flexibility to the user to specify how parameters should be drawn as any function can be used and any parameter connections are accepted. 

The \texttt{skysurvey} package contains three main parts. The first is \texttt{Target} that specifies how the astrophysical population exists in nature. The \texttt{Target} parameters are based on \texttt{modeldag} and the spectro-photometric (time series) template that provides the target flux as a function of time, wavelength or filter are based on \texttt{sncosmo}. As such, any \texttt{sncosmo} template or model (built-in or user-defined) are accepted by \texttt{skysurvey}. By default \texttt{skysurvey} has many built-in \texttt{Target} such as any Supernovae type (SNeII, SNeIa, SNeIbc etc.) or kilonovae. The second part is \texttt{Survey}, that carries the information of how the sky has been observed. A user can easily build any customed survey (see online documentation) or use a list of predefined ones such as \texttt{ZTF} or \texttt{LSST}. the third part is \texttt{DataSet}, which builds the expected observations of \texttt{Target} (or list of) by the \texttt{Survey}. The user can focus on complexifying the astrophysical population at will, they may alter the \texttt{Survey} -- for instance to introduce calibration uncertainties, but the \texttt{DataSet} will still be able to handle them. That is the strength of \texttt{skysurvey}: the package is intuitive, user oriented, and designed to be highly robust and easy to use. It is finally solely based on well supported python libraries such as \texttt{numpy}, \texttt{pandas} and \texttt{sncosmo}, ensuring stability and long term user supports and maintenance.

\begin{acknowledgements}
This paper has not been written with the support of any large language models. This project has received funding from the European Research Council (ERC) under the European Union's Horizon 2020 research and innovation program (grant agreement n 759194 - USNAC. M.W.C. acknowledges support from the National Science Foundation with grant numbers PHY-2117997, PHY-2308862 and PHY-2409481.
\end{acknowledgements}

\end{document}